\documentclass[%
 reprint,amsmath,amssymb,aps,prl,longbibliography,preprintnumbers,showpacs,floatfix,
nofootinbib
]{revtex4-1}

\usepackage{color,graphicx}
\usepackage{bm}

\newcommand{\ee}{e^+ e^-}
\newcommand{\pp}{\pi^+ \pi^-}
\newcommand{\kk}{K^+ K^-}
\newcommand{\ep}{\eta \pi^0}
\newcommand{\upp}{\Upsilon(1S) \pi^+ \pi^-}
\newcommand{\ukk}{\Upsilon(1S) K^+ K^-}
\newcommand{\uep}{\Upsilon(1S) \eta \pi^0}
\newcommand{\uos}{\Upsilon(1S)}
\newcommand{\uts}{\Upsilon(2S)}
\newcommand{\uths}{\Upsilon(3S)}

\newcommand{\yb}{Y_{b}}

\begin{document}

\preprint{DESY 10-076}

\title{Tetraquark-based analysis and predictions of the cross sections
  and distributions\\ for the processes 
  $e^+ e^- \to \Upsilon(1S)(\pi^+\pi^-, K^+ K^-, \eta\pi^0)$ 
  near $\Upsilon(5S)$}

\author{Ahmed~Ali}
\email{ahmed.ali@desy.de}
\author{Christian~Hambrock}
\email{christian.hambrock@desy.de}
\author{Satoshi~Mishima}
\email{satoshi.mishima@desy.de}
\affiliation{Deutsches Elektronen-Synchrotron DESY, D-22607 Hamburg, Germany}

\date{\today}

\begin{abstract}
We calculate the cross sections and final state distributions for the
processes $e^+ e^- \to \Upsilon(1S)(\pi^+\pi^-, K^+ K^-, \eta\pi^0)$
near the $\Upsilon(5S)$ resonance based on the tetraquark hypothesis. 
This framework is used to analyse the 
data on the 
$\Upsilon(1S)\pi^+\pi^-$ and $\Upsilon(1S)K^+ K^-$ final states
[K.~F.~Chen {\it et al.} (Belle Collaboration), Phys.\ Rev.\ Lett.\
{\bf 100}, 112001 (2008); I.~Adachi {\it et al.} (Belle Collaboration), 
arXiv:0808.2445], yielding good fits. Dimeson invariant mass spectra
in these processes are shown to be dominated by the corresponding
light scalar and tensor states. The resulting correlations among the
cross sections are worked out. We also predict $\sigma(e^+ e^- \to
\Upsilon(1S) K^+K^-)/\sigma(e^+ e^- \to \Upsilon(1S) K^0
\bar{K}^0)=1/4$. 
These features provide crucial tests of the tetraquark
framework and can be 
searched for in the currently available and forthcoming data from the $B$ factories. 
\end{abstract}

\pacs{14.40.Rt, 14.40.Pq, 13.66.Bc}

\maketitle

%
%

The anomalously large production cross sections for $e^+e^- \to \uos
\pp, \uts \pp$ and $e^+e^- \to \uths \pp$ measured between
$\sqrt{s}=10.83$ GeV and 11.02 GeV by the Belle
Collaboration~\cite{Abe:2007tk,:2008pu} at KEK do not agree well with
the lineshape and production rates for the conventional $b\bar{b}$
state $\Upsilon(10860)$ (also called $\Upsilon(5S)$). A fit to the
measured production cross sections using a Breit-Wigner resonance
shape yielded a peak mass of  
$[10888.4^{+2.7}_{-2.6} ({\text{stat}})\pm 1.2({\text{syst}})]$ MeV
and a width of 
$[30.7 ^{+8.3}_{-7.0}({\text{stat}}) \pm 3.1({\text{syst}})]$ MeV for
the observed state, henceforth called
$Y_b(10890)$~\cite{:2008pu}. More data are required to understand the
resonance structure in this region. In~\cite{Ali:2009es}, a dynamical
model was developed to explain the Belle data  for the final states
$\uos \pp$ and $\uts \pp$ in terms of the production and decays of the
states $Y_{[b,l/h]}$, which are linear superposition of the
$J^{PC}=1^{--}$ hidden $b\bar{b}$ tetraquark states 
$Y_{[bu]}\equiv[bu][\bar{b}\bar{u}]$ and
$Y_{[bd]}\equiv[bd][\bar{b}\bar{d}]$. The mass difference, estimated
as $M(Y_{[b,h]})- M(Y_{[b,l]})= (5.6 \pm 2.8)$
MeV~\cite{Ali:2009pi,Maiani:2004vq}, was ignored and the
mass-degenerate states $Y_{[b,l]}$ and $Y_{[b,h]}$ were identified 
with the $Y_b(10890)$. This model described the distributions in
the dipion invariant mass and the helicity angle measured by
Belle~\cite{Abe:2007tk} well and offered an explanation of the
rates in terms of the Zweig-allowed transitions $Y_{[b,l/h]}\to (\uos,
\uts, \uths) \pp$. 
In the case of $X(3872)$, the phase space for the decay $X(3872) \to 
D D^{*}$ is highly constrained, yielding a very small binding energy:
${\cal E}_0 \simeq M_X - M_D - M_{D^*}= -0.25 \pm 0.40$ MeV. This led
to suggestions that $X(3872)$ could be a hadron molecule. In the case
of $Y_b(10890)$, the phase space is large enough to allow the decays
$Y_b(10890) \to B^{(*)} \bar{B}^{(*)}$, and consequently the decay
width is large, $\Gamma_{Y_b} \simeq 30$ MeV, which is of the same
order as the total width of the $\Upsilon(5S)$. Hence, $Y_b(10890)$,
in all likelihood, is not a hadron molecule. 
While credible, the interpretation of $Y_b(10890)$
in terms of the $b\bar{b}$ tetraquark states 
requires further experimental scrutiny. It is the aim of
this Letter to provide some definitive tests to confirm or rule out
the tetraquark interpretation of the Belle data.

To that end, we further develop the tetraquark formalism for the
processes 
\begin{align}
e^+ + e^-\to Y_b(q)\to \uos(p) + P(k_1) + P^\prime(k_2)\,,
\label{eq:process}
\end{align}
where $P P'$ stands for the pseudoscalar-meson pairs $\pp$, $\kk$ and
$\ep$, and $q$, $p$, $k_1$ and $k_2$ are the momenta of $Y_b$, $\uos$,
$P$ and $P'$, respectively. We neglect other background processes,
based on prior data on the dipionic transitions involving higher
$\Upsilon(nS)$ to lower $\Upsilon(mS)$ ($m<n$)
states~\cite{Amsler:2008zzb}. Following~\cite{Hooft:2008we}, the low
mass scalar $0^{++}$ hadrons $\sigma$ or $f_0(600)$, $f_0(980)$ and
$a_0^0(980)$ (the upper index indicates the $I_3=0$ component of the
iso-triplet $a_0$), which enter as intermediate states in the processes 
$Y_b \to \Upsilon(1S) + [\sigma,\, f_0(980),\,\cdots] \to 
\Upsilon(1S)  P P^\prime$,
are assumed to be tetraquark states. These intermediate
$J^{\text{PC}}=0^{++}$ states together with the $J^{\text{PC}}=2^{++}$
state $f_2(1270)$ provide the dominant resonating part of the
amplitudes for the processes considered in this work. These resonances
are labeled $\sigma$, $f_0$, $a_0^0$ and $f_2$ henceforth. We
determine the coupling constants involving these light tetraquark
states and the mesons $PP'$ from the known decays from the
PDG~\cite{Amsler:2008zzb} and data from the E791~\cite{Aitala:2000xu}, 
the BES~\cite{Ablikim:2004wn}, the Crystal Barrel (CB)~\cite{Abele:1998qd}
and the KLOE Collaborations~\cite{Ambrosino:2009zzb,Ambrosino:2009py},
adopting the Flatt\'e model~\cite{Flatte:1976xu} for the $\sigma$,
$f_0$ and $a_0^0$ couplings to take into account threshold effects. 
The non-resonating continuum contributions are parameterized in terms
of two {\it a priori} unknown constants~\cite{Brown:1975dz}. 
With this formalism, we analyze the invariant-mass $M_{PP'}$ and the
$\cos \theta$ spectra, where $M_{PP'}=(k_1+k_2)^2$ and $\theta$ is the
angle between the momenta of $\yb$ and $P$ in the $PP'$ rest frame. 
 
The theoretical framework described here provides good fits of the
Belle data on the invariant dipion mass spectrum and $\cos \theta$
distribution in the process $e^+ e^- \to \Upsilon(1S) \pi^+\pi^-$ and
the ratio $\sigma_{\uos\kk}/\sigma_{\uos\pp}$, with $\sigma_{\uos PP'}$ 
being the cross section $\sigma( e^+e^- \to \Upsilon(1S)P P')$. 
We present the invariant mass distributions for the $\kk$ and $\ep$
mesons in the processes $e^+ e^- \to \Upsilon(1S)(K^+ K^-,
\eta\pi^0)$, which are dominated by the respective 
$J^{PC}=0^{++}$ resonances. The resulting correlations among
$\sigma_{\Upsilon(1S) \pi^+\pi^-}$, $\sigma_{\Upsilon(1S) K^+K^-}$ and
$\sigma_{\Upsilon(1S) \eta \pi^0}$ are worked out. Constraining these
correlations from the existing data on the first two processes, we
predict $\sigma_{\uos\ep}/\sigma_{\uos\pp}$. We also predict
$\sigma_{\uos\kk}/\sigma_{\uos K^0 \bar{K}^0}=1/4$, reflecting the
ratio $Q_{[bu]}^2/Q_{[bd]}^2$  with $Q_{[bu]}=1/3$ and $Q_{[bd]}=-2/3$
being the effective electric charges for the constituent diquarks of
$Y_{[bu]}$ and $Y_{[bd]}$, respectively.

%
%

We start by defining the tetraquark states in the isospin basis, 
with the two isospin components 
$\yb ^0 \equiv (Y_{[bu]}+Y_{[bd]})/\sqrt{2}$ and
$\yb ^1 \equiv (Y_{[bu]}-Y_{[bd]})/\sqrt{2}$
for isospin $I=0$ and $I=1$, respectively. The two mass eigenstates
$Y_{[b,l]}$ and $Y_{[b,h]}$ are identified with $Y_{[bu]}$ and
$Y_{[bd]}$, as the mixings between them is small. We ignore the mass
difference and also the isospin breaking effects except for the
production processes $\ee\to \yb^I$ hereafter.  

We calculate the decay amplitude as a sum of the Breit-Wigner
resonances and non-resonating continuum contributions, with the
latter adopted from \cite{Brown:1975dz}. The differential cross
section is then written as 
\begin{align}
\frac{d^2\sigma_{\uos PP'}}{dM_{PP'}\, d\cos\theta}
&=
\frac{\lambda^{1/2}(s, m_\Upsilon^2, M_{PP'}^2)
\lambda^{1/2}(M_{PP'}^2, m_P^2, m_{P'}^2)}
{384\pi^3 s\, M_{PP'}
\left[ (s-m_{Y_b}^2)^2+m_{Y_b}^2\Gamma_{Y_b}^2 \right]}
\nonumber\\
&\hspace{-18mm}\times
\Bigg\{
\left( 1  + \frac{(q\cdot p)^2}{2s\, m_\Upsilon^2} \right)
\left| {\cal S} \right|^2
\nonumber\\
&\hspace{-13mm}
+
2\, \text{Re} \left[
{\cal S}^*
\left(
{\cal D}'
+  \frac{(q\cdot p)^2}{2s\, m_\Upsilon^2}\,
{\cal D}''
\right)
\right]
\left( \cos^2\theta - \frac{1}{3} \right)
\nonumber\\
&\hspace{-13mm}
+
\left| {\cal D} \right|^2
\sin^2\theta
\left[
 \sin^2\theta
 + 2\left(\frac{(q^0)^2}{s}\,
 + \frac{(p^0)^2}{m_\Upsilon^2} \right)
 \cos^2\theta
\right]
\nonumber\\
&\hspace{-13mm}
+
\left(
\left| {\cal D}' \right|^2
+
\frac{(q\cdot p)^2}{2s\, m_\Upsilon^2}
\left| {\cal D}'' \right|^2
\right)
\left( \cos^2\theta - \frac{1}{3} \right)^2
\Bigg\}\,,
\end{align}
where $\lambda(x,y,z)\equiv(x-y-z)^2-4yz$, $q^0$ and $p^0$ are the
energies of the $Y_b$ and $\uos$ in the $PP'$ rest frame, respectively, 
$\Gamma_{Y_b}$ is the decay width of $Y_b$, and $m_{Y_b}$, $m_{\Upsilon}$,
$m_P$ and $m_{P'}$ are the masses of $Y_b$, $\uos$, $P$ and $P'$,
respectively. We take $m_{Y_b}=10.89$ GeV and $\Gamma_{Y_b}=30$ MeV in the
numerical analysis below. A detailed derivation of the above formula
will be presented in \cite{Ali:longpaper}.   

Each $PP'$ channel receives specified contributions depending on the
isospin of $PP'$ and the kinematically allowed region for the
invariant mass $M_{PP'}\in [m_{P}+m_{P'}, \sqrt{s}-m_{\Upsilon(1S)}]$. 
The S-wave amplitude for the $PP'$ system, ${\cal S}$, and the D-wave
amplitudes, ${\cal D}$, ${\cal D}'$ and ${\cal D}''$, are the sums
over possible isospin states 
\begin{align}
&{\cal M}  = \sum_{I} {\cal M}_{I}
\ \ \ \text{for}\ \ {\cal M}={\cal S},\ {\cal D},\
{\cal D}',\ {\cal D}'',
\end{align}
where $I=0$ for $\pp$, $I=0,1$ for $\kk$, and $I=1$ for $\ep$, since
the $\uos$ is an isospin 0 state, and the following resonances
contribute to each process: 
\begin{align}
\begin{array}{ll}
\sigma,\ f_0\ \, \text{and}\ \, f_2 & \text{for}\ \ \pp, \\
f_0,\  a_0^{0}\ \, \text{and}\ \, f_2 & \text{for}\ \ K^+K^-, \\
a_0^{0} & \text{for}\ \ \eta\pi^0.
\end{array}
\label{eq:resonances}
\end{align}
The $I=0$ amplitudes are given by the combinations of the resonance
amplitudes, ${\cal M}_0^S$ and ${\cal M}_0^{f_2}$, and the
non-resonating continuum amplitudes, ${\cal M}_{0}^{1C}$ and 
${\cal M}_{0}^{2C}$: 
\begin{align}
&{\cal S}_{0} =
{\cal M}_{0}^{1C} + (k_1\cdot k_2) \sum_{S} {\cal M}_0^S,
\ \ \
{\cal D}_{0} = |\bm{k}|^2 {\cal M}_0^{f_2},
\nonumber\\
&{\cal D}_0' = {\cal M}_{0}^{2C} - {\cal D}_{0}\,,
\ \ \
{\cal D}_0'' =
{\cal M}_{0}^{2C} + \frac{2q^0p^0}{(q\cdot p)}
{\cal D}_{0}\,,
\label{eq:izerodef}
\end{align}
where $S$ runs over possible $I=0$ scalar resonances in
Eq.~\eqref{eq:resonances}, and $|\bm{k}|$ is the magnitude of the
three momentum of $P^{(\prime)}$ in the $PP'$ rest frame.
Similarly, the $I=1$ amplitudes are given by
\begin{align}
&{\cal S}_{1} =
\frac{g_{e^+e^-Y_b^1}}{g_{e^+e^-Y_b^0}}
\left[{\cal M}_{1}^{1C} + (k_1\cdot k_2) {\cal M}_1^{a_0^0}\right],
\nonumber\\
&{\cal D}_{1} = 0\,,
\ \ \
{\cal D}_{1}' =
{\cal D}_{1}'' =
\frac{g_{e^+e^-Y_b^1}}{g_{e^+e^-Y_b^0}}{\cal M}_{1}^{2C},
\label{eq:ionedef}
\end{align}
where the dimensionless couplings $g_{e^+e^-Y_b^{I}}$ are defined
through the Lagrangian
${\cal L}
= \sum_{I=0,1} g_{e^+e^-Y_b^I} Y_{b\mu}^I\, (\bar{e}\gamma^\mu e)$, 
and the ratio is given by 
$g_{e^+e^-Y_b^1}/g_{e^+e^-Y_b^0} 
= ( Q_{[bu]} -Q_{[bd]} )/ ( Q_{[bu]} +Q_{[bd]} )= -3$. 

To calculate the production cross sections, we derive the
corresponding Van Royen-Weisskopf formula for the leptonic decay
widths of the tetraquark states made up of point-like diquarks:
\begin{align}
\Gamma( Y_{[bu/bd]} \to e^+e^-)
=
\frac{24\alpha^2|Q_{[bu/bd]}|^2}{m_{Y_b}^4}\,
\kappa^2\left|R^{(1)}_{11}(0)\right|^2,
\label{eq:VRW-P}
\end{align}
where $\alpha$ is the fine-structure constant, the parameter $\kappa$
takes into account differing sizes of the tetraquarks compared to the
standard bottomonia as well as the hadronic size of the diquarks, 
with $\kappa < 1$ anticipated, 
and $| R^{(1)}_{11}(0)|^2=2.067$ GeV$^5$~\cite{Eichten:1995ch}
is the square of the derivative of the radial wave function for
$\chi_b(1P)$ taken at the origin. Hence, the leptonic widths of the
tetraquark states are estimated as  
\begin{align}
\Gamma( Y_{[bd]} \to e^+e^-) 
= 4\,\Gamma( Y_{[bu]} \to e^+e^-) 
\approx 83\, \kappa^2\ \text{eV}\,,
\label{eq:Ybtoee_value}
\end{align}
which are substantially smaller than the leptonic width of the
$\Upsilon(5S)$~\cite{Amsler:2008zzb}. 
Combining the knowledge of the production
process and the measured cross section for $\ee\to\uos\pp$, we estimate
$\Gamma(Y_{[bu/bd]}\to\upp)$ to be $O(1)$ MeV.

The continuum amplitudes in Eq.~\eqref{eq:izerodef} are written in
terms of the two form factors $A$ and
$B$~\cite{Brown:1975dz,Ali:longpaper} as 
\begin{align}
{\cal M}_{0}^{1C}
&=
\frac{2A'}{f_Pf_{P'}}(k_1\cdot k_2)
+ \frac{B'}{f_Pf_{P'}}
\frac{3(q^0)^2k_1^0k_2^0-|\bm{q}|^2|\bm{k}|^2}{3s}
\,,\nonumber\\
{\cal M}_{0}^{2C}
&=
- \frac{B'}{f_Pf_{P'}}
\frac{|\bm{q}|^2|\bm{k}|^2}{s}
\,,
\label{eq:amp_decay-cont}
\end{align}
where the primed quantities here, and later, are rescaled as
$A^\prime =A g_{e^+e^-Y_b^0}$ and $B^\prime =B g_{e^+e^-Y_b^0}$, 
$f_{P^{(\prime)}}$ is the decay constant of $P^{(\prime)}$, and
$|\bm{q}|$, $k_1^0$ and $k_2^0$ are the magnitude of the three
momentum of $Y_b$ and the energies of $P$ and $P'$ in the $PP'$ rest
frame, respectively. Using SU(3) symmetry for the form factors $A$
and $B$ in Eq.~\eqref{eq:amp_decay-cont}, but not for the pseudoscalar
meson masses and coupling constants, we assume the relations
${\cal M}_{0}^{1C,2C}(\ukk) = (\sqrt{3}/2)\, {\cal M}_{0}^{1C,2C}(\upp)$, 
${\cal M}_{1}^{1C,2C}(\ukk) = {\cal M}_{0}^{1C,2C}(\ukk)$
and ${\cal M}_{1}^{1C,2C}(\uep) = \sqrt{2}\, {\cal M}_{1}^{1C,2C}(\ukk)$.

The resonant contributions are expressed by the Breit-Wigner formula:
\begin{align}
\mathcal{M}^{R}_I
&=
\frac{g_{R PP'}\, g_{\yb^I \Upsilon(1S) R}\, g_{\ee Y_b^0}}
{M_{PP'}^2 -m_{R}^2+ i\, m_{R}  \Gamma_{R}}\, e^{i\varphi_{R}},
\label{eq:amp_decay}
\end{align}
where $I=0$ for $R=\sigma$, $f_0$ and $f_2$, and $I=1$ for $R=a_0^0$.
The couplings for the scalar resonances $S$ are defined through the
Lagrangian  
${\cal L} = g_{SPP'} (\partial_\mu P) (\partial^\mu P')\,S
+ g_{Y_b \Upsilon(1S) S}\,Y_{b\mu} \Upsilon^\mu S$,
while those for 
the $f_2$ 
are defined via 
${\cal L} = 2\hspace{0.3mm} g_{f_2PP'}
(\partial_\mu P)(\partial_\nu P') f_2^{\mu\nu}
+ g_{Y_b \Upsilon(1S) f_2}Y_{b\mu} \Upsilon_\nu f_2^{\mu\nu}$.
The couplings $g_{R PP'}$ and $g_{\yb^I \Upsilon(1S) R}$ have mass
dimensions $-1$ and $1$, respectively. 
For the $\sigma$, $f_0$ and $a_0^0$, we adopt the Flatt\'e
model~\cite{Flatte:1976xu} 
\begin{align}
&m_{\sigma}\Gamma_{\sigma} = f_{\sigma\pi\pi}^2 \rho_{\pi\pi}\,,
\ \
m_{f_0}\Gamma_{f_0} =
f_{f_0\pi\pi}^2 \rho_{\pi\pi} + f_{f_0K\bar{K}}^2 \rho_{K\bar{K}}\,,
\nonumber\\
&m_{a_0^0}\Gamma_{a_0^0} =
f_{a_0^0\eta\pi}^2 \rho_{\eta\pi} + f_{a_0^0K\bar{K}}^2 \rho_{K\bar{K}}
\end{align}
with the phase space factor
$\rho_{ab}=[(1-(m_a-m_b)^2/M_{PP'}^2)(1-(m_a+m_b)^2/M_{PP'}^2)]^{1/2}$,
where the Flatt\'e couplings $f_{S PP'}$ are related to the vertex
couplings $g_{SPP'}$ entering in Eq.~(\ref{eq:amp_decay}) via
\begin{align}
g_{S PP'}(k_1\cdot k_2)
= 4\sqrt{ \pi}\, f_{S PP'}.
\label{eq:couplings_scalars}
\end{align}
The couplings in Eq.~\eqref{eq:couplings_scalars} are defined for an
exclusive final state. Summing over final states, {\it e.g.}, 
$f_{f_0 \pi\pi}^2=f_{f_0 \pp}^2 + f_{f_0 \pi^0\pi^0}^2$, we obtain
the isospin relations
$f_{S \pi\pi } = \sqrt{3/2}\, f_{S \pp}$,
$f_{S K\bar{K}} = \sqrt{2}\, f_{S \kk}$ and
$f_{S \eta\pi} = f_{S \eta\pi^0}$.
For the $\sigma$ meson, we extract the coupling $g_{\sigma\pp}$ from
the E791 data~\cite{Aitala:2000xu}:
$g_{\sigma\pp} = 26.7\ \text{GeV}^{-1}$ with $m_{\sigma}=478$ MeV,
yielding the Flatt\'e coupling $f_{\sigma\pi\pi}=437$ MeV. For the
$f_0$ and $a_0^0$ mesons, we adopt the masses and the Flatt\'e
couplings measured by the BES~\cite{Ablikim:2004wn} and
CB~\cite{Abele:1998qd} Collaborations (the corresponding couplings
from KLOE~\cite{Ambrosino:2009zzb,Ambrosino:2009py} are shown in the 
parentheses): 
\begin{align}
&\!\!\!
m_{f_0}\! =\! 965 (984),\,  
f_{f_0\pi\pi}\! =\! 406 (349),\,  
f_{f_0 K \bar{K}}\! =\! 833 (869),
\nonumber\\
&\!\!\!
m_{a_0^0}\! =\! 982 (983),\,
f_{a_0^0 \eta\pi}\! =\! 324 (398),\,  
f_{a_0^0 K \bar{K}}\! =\! 329 (429)\!
\label{eq:CBBES-KLOEinputs}
\end{align}
in units of MeV. Furthermore, we extract the couplings for the $f_2$
meson from  
$\Gamma(f_2\to PP')=
g_{f_2 PP'}^2 m_{f_2}^3( 1 - 4m_P^2/m_{f_2}^2)^{5/2}/(480\pi)$
for $m_P=m_{P'}$, where the data for 
$\Gamma(f_2\to\pi\pi) = (3/2)\,\Gamma(f_2\to\pp)$ and 
$\Gamma(f_2\to K\bar{K}) = 2\,\Gamma(f_2\to\kk)$, and 
$m_{f_2}=1275$ MeV are taken from PDG~\cite{Amsler:2008zzb}. 
The other inputs for the pseudo-scalar mesons and the $\Upsilon(1S)$
are also taken from PDG.

%
%

Having detailed our dynamical model, we now perform a simultaneous fit
to the binned $\upp$ data for the $M_{\pp}$ and $\cos\theta$
distributions measured by Belle at $\sqrt{s}=10.87$
GeV~\cite{Abe:2007tk}, normalizing them by the measured cross section: 
$d\widetilde{\sigma}_{\pp}/dM_{\pp}$ and 
$d\widetilde{\sigma}_{\pp}/d\cos\theta$, where
$\widetilde{\sigma}_{\pp}\equiv\sigma_{\uos\pp}/\sigma_{\uos\pp}^{\text{Belle}}$
with $\sigma_{\uos \pp}^{\text{Belle}}=1.61 \pm 0.16$ pb~\cite{Abe:2007tk}. 
With SU(3) symmetry for the $Y_b^0 \uos R$ couplings entering in
Eq.~(\ref{eq:amp_decay}), {\it i.e.}, setting 
$g_{\yb^0 \Upsilon(1S) \sigma}=g_{\yb^0 \Upsilon(1S) f_0}$, 
we have 7 free parameters: 
\begin{align}
A',\
B',\
g_{\yb^0 \Upsilon(1S) f_0}',\
g_{\yb^0 \Upsilon(1S) f_2}',\
\varphi_\sigma,\
\varphi_{f_0},\
\varphi_{f_2}.
\label{eq:fitting_parameters}
\end{align}
Hence the number of degrees of freedom
(d.o.f.) in the fit is $( 12 + 10 ) - 7 = 15$.
To make predictions for the $\uos\kk$ and $\uos\ep$ modes, we assume
SU(3) for the couplings and the phases, {\it i.e.}, 
$g_{\yb^0 \Upsilon(1S) f_0} = g_{\yb^1 \Upsilon(1S) a_0^0}$ and
$\varphi_{f_0} = \varphi_{a_0^0}$. We also assume that there is no phase
difference between the two continuum amplitudes 
${\cal M}^{1C,2C}_{0}$ and ${\cal M}^{1C,2C}_{1}$.

\begin{figure}[t]
\includegraphics[width=0.235\textwidth]{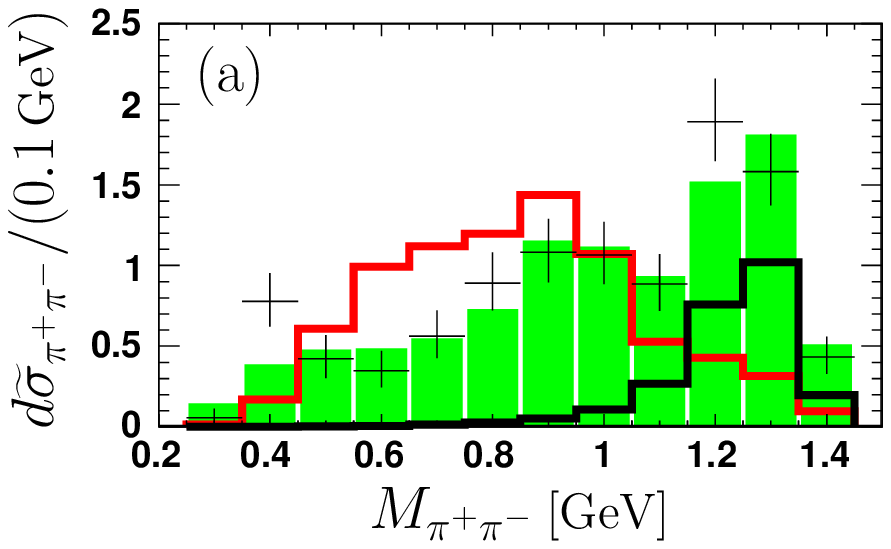}
\includegraphics[width=0.235\textwidth]{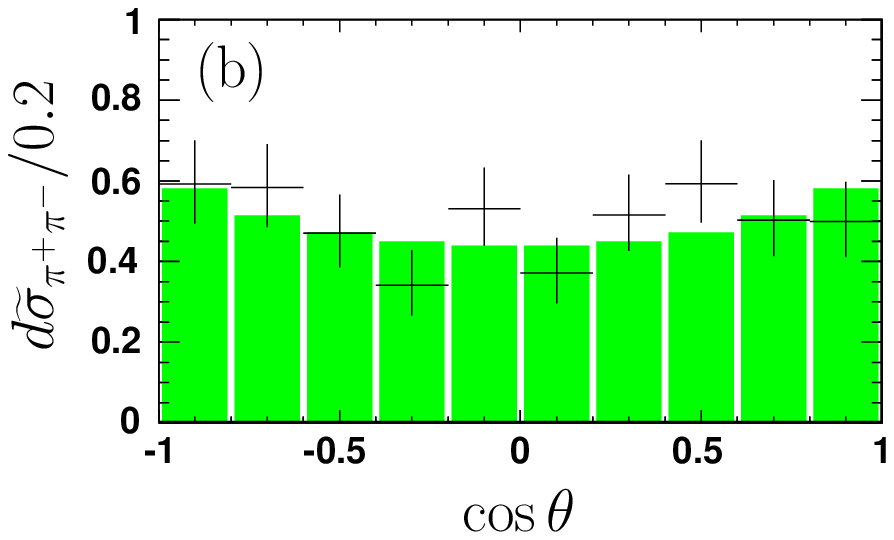}
\caption{Fit results (a) of the $M_{\pp}$ distribution and (b) of the
  $\cos\theta$ distribution for $\ee\to Y_b \to \upp$, 
  normalized by the measured cross section. The histograms (green
  bars) represent the fit results, while the crosses are the 
  Belle data~\cite{Abe:2007tk}. The resonant contributions from the
  $\sigma$ and $f_0(980)$ (left red curve) and the $f_2(1270)$ (right
  black curve) are also indicated in (a). 
\label{fig:spectra}
}
\end{figure}
\begin{table}[t]
\caption{Best fit parameters, yielding $\chi^2/{\text{d.o.f.}} =
  21.5/15$, where $A'$ and $B'$ are dimensionless, $g_{\yb^0
    \Upsilon(1S) f_0}'$ and $g_{\yb^0 \Upsilon(1S) f_2}'$ are given in
  units of MeV, and the angles are in units of rad. 
\label{fitvaluest1res1s}
}
\begin{ruledtabular}
\begin{tabular}{ccccccc}
 $A'$ & $B'$ &
 $g_{\yb^0 \Upsilon(1S) f_0}'$ &
 $g_{\yb^0 \Upsilon(1S) f_2}'$ &
 $\varphi_\sigma$ & $\varphi_{f_0}$ & $\varphi_{f_2}$
\\
\hline
 $0.000079$ & $-0.00020$ &
 $0.318$ &  $0.439$ &
 $0.36$ & $-2.76$ & $-0.46$
\\
\end{tabular}
\end{ruledtabular}
\end{table}

\begin{figure*}[!t]
\includegraphics[width=0.3\textwidth]{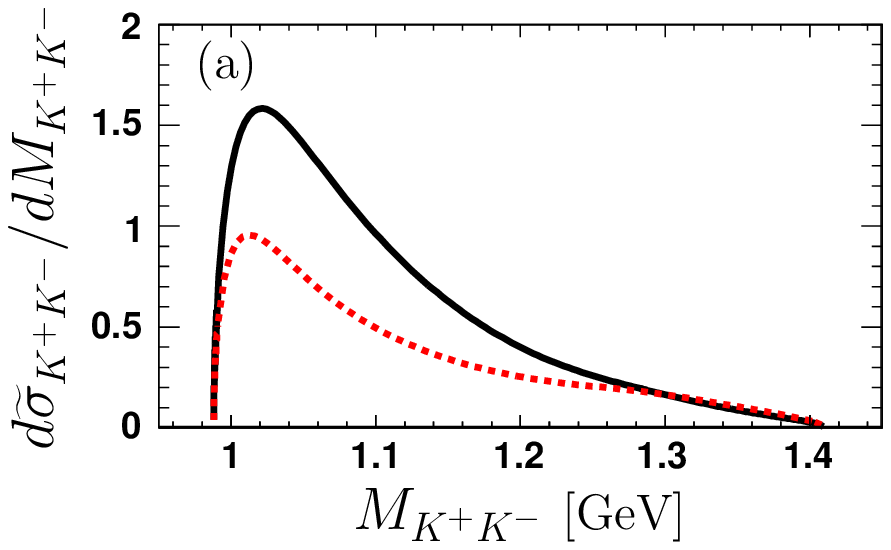}\hspace{2mm}
\includegraphics[width=0.3\textwidth]{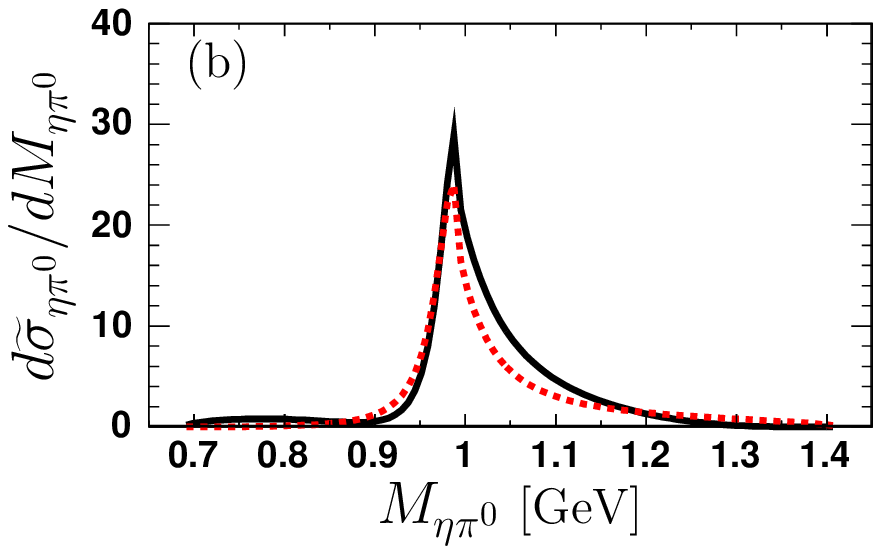}\hspace{2mm}
\includegraphics[width=0.3\textwidth]{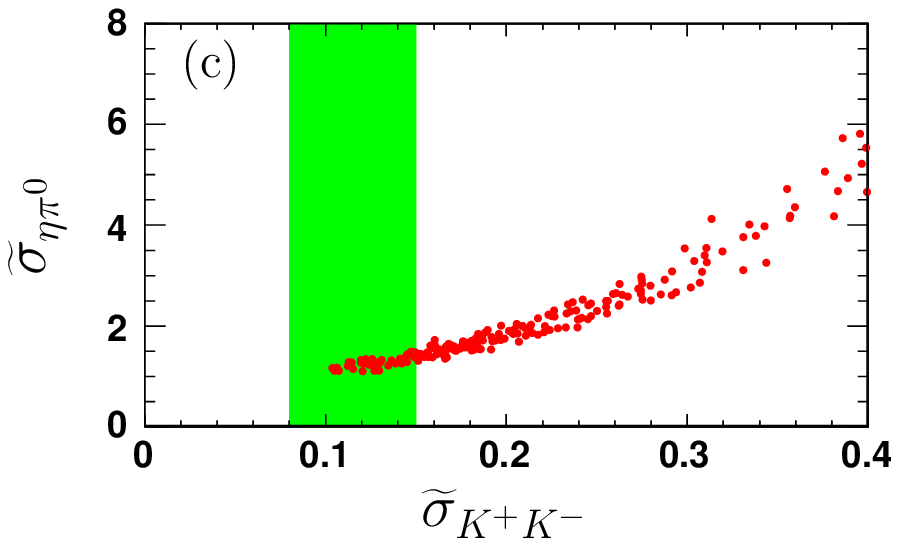}
\caption{Predictions (a) of the $M_{\kk}$ distribution for
  $\ee\to Y_b \to \ukk$, (b) of the $M_{\ep}$ distribution for
  $\ee\to Y_b \to \uep$ and (c) of the correlation between the cross
  sections of $\ukk$ and $\uep$, normalized by the measured cross
  section for the $\upp$ mode. In (a) and (b), the dotted (solid)
  curves show the dimeson invariant mass spectra from the resonant
  (total) contribution. In (c), the red dots represent predictions
  from our fit solutions satisfying $\chi^2/{\text{d.o.f.}}< 1.6$. 
  The shaded (green) band shows the current Belle measurement 
  $\widetilde{\sigma}_{K^+K^-}= 0.11^{+0.04}_{-0.03}$~\cite{Abe:2007tk}.  
\label{fig:predictions}
}
\end{figure*}

With these inputs, we have performed a large number of fits (typically
$O(5000)$) of the Belle data with the tetraquark theory predictions. 
The resultant best fit is fairly good, with $\chi^2/{\text{d.o.f.}} =
21.5/15$ for the BES and CB input in Eq.~\eqref{eq:CBBES-KLOEinputs}, 
which corresponds to a $p$-value of 0.12. The corresponding best fit
using the KLOE data is very similar, having a $\chi^2/{\text{d.o.f.}}
= 21.9/15$, yielding a $p$-value of 0.11. The best fit using the BES
and CB data is presented in Fig.~\ref{fig:spectra} and the
corresponding fit values of the parameters are listed in
Table~\ref{fitvaluest1res1s}. Further details about the correlations
among the parameters and the cross sections will be presented in a
forthcoming paper~\cite{Ali:longpaper}. Concluding the discussion of
the final state $\uos \pp$, we note that the resonance contribution
represented by the left red curve (S-wave from $\sigma$ and $f_0$) and
the right black curve (D-wave from $f_2$) in Fig.~\ref{fig:spectra}
(a) dominate the $M_{\pp}$ spectrum, supporting our dynamical model in
the decay $Y_b \to \uos \pp$. Sufficient data may provide enough
statistics to undertake an analysis in the end-region of $M_{\pp}$ to
probe the angular distribution of $f_2 \to \pi^+\pi^-$. 

The normalized $M_{K^+K^-}$ and $M_{\eta \pi^0}$ distributions,
calculated with the best-fit parameters in
Table~\ref{fitvaluest1res1s}, are shown in Fig.~\ref{fig:predictions}
(a) and Fig.~\ref{fig:predictions} (b), respectively. In these
figures, the dotted (solid) curves show the dimeson invariant mass
spectra from the resonant (total) contribution. Since these spectra
are dominated by the scalars $f_0+a_0^0$ and $a_0^0$, respectively,
there is a strong correlation between the two cross sections. This is
shown in Fig.~\ref{fig:predictions} (c), where we have plotted the
normalized cross sections $\widetilde{\sigma}_{K^+K^-}$ and
$\widetilde{\sigma}_{\eta \pi^0}$ resulting from our fits (dotted 
points) which all satisfy $\chi^2/{\text{d.o.f.}} < 1.6$.
The current Belle measurement
$\widetilde{\sigma}_{K^+K^-}=0.11^{+0.04}_{-0.03}$~\cite{Abe:2007tk}
is shown as a shaded (green) band on this figure. Our model is in
agreement with the Belle measurement, though there is a tendency in
the fits to yield larger value for  $\widetilde{\sigma}_{K^+K^-}$.  
Our predictions will be further tested as and when the cross section
$\widetilde{\sigma}_{\eta \pi^0}$ is measured. Noticing that we have
neglected the SU(3)-breaking effects, we predict 
$1.0 \lesssim \widetilde{\sigma}_{\eta \pi^0} \lesssim 2.0$.  

Finally, we note that the states $\Upsilon(1S)\kk$ and 
$\Upsilon(1S) K^0\bar{K}^0$ are produced by the underlying mechanism 
$e^+e^-\to Y_{[bu]}\to \Upsilon(1S)\kk$ and 
$e^+e^-\to Y_{[bd]}\to \Upsilon(1S) K^0\bar{K}^0$. 
Hence, a firm prediction is 
\begin{align}
\frac{\sigma_{\Upsilon(1S)\kk}}
{\sigma_{\Upsilon(1S) K^0 \bar{K}^0}}
=
\frac{Q_{[bu]}^2}{Q_{[bd]}^2}
=
\frac{1}{4}\,. 
\end{align}
This relation is valid under the assumption that the diquarks are
point-like. In terms of the mass eigenstates, we predict
$\sigma_{\Upsilon(1S)\kk} = \sigma_{\Upsilon(1S) K_S K_S}$. 

The distributions, cross sections, and correlations presented here are
crucial tests of the underlying tetraquark hypothesis in the $b\bar{b}$ 
sector and go well beyond what has been proposed in the literature to 
understand the nature of the $Y_b(10890)$ state~\cite{Ali:2009es}. 
They will be scrutinized soon in the existing and the forthcoming
Belle data to which we look forward.

\end{document}